\documentclass[twocolumn,showpacs,preprintnumbers,amsmath,amssymb,aps,prl]{revtex4}

\usepackage{graphicx}
\usepackage{dcolumn}
\usepackage{bm}
\begin{document}

\preprint{LA-UR 04-3937}

\title{New theoretical description of neutron scattering in a monatomic liquid}

\author{Duane C. Wallace}
\author{Giulia De Lorenzi-Venneri}
\affiliation{Theoretical Division, Los Alamos National Laboratory, 
Los Alamos, New Mexico 87545}

\date{\today}

\begin{abstract}
In a recently developed theory of the atomic motion in monatomic liquids, the motion is comprised of normal mode vibrations
in any of the large number of equivalent random valleys, interspersed with nearly instantaneous transits 
which carry the system between neighboring valleys. The consequences for $S(q,\omega )$ are presented here:
when the system moves in a single random valley, the inelastic part of  $S(q,\omega )$ is a sum over all vibrational
modes of the inelastic cross section of each single mode; in the liquid state the system undergoes transits
at a rapid rate, causing the Rayleigh and Brillouin peaks to broaden but not to shift; over the entire $q$ range where 
the Brillouin peak is distinguishable, its location in the liquid is the same as it is in a single random valley. These 
properties are verified by comparison between theory and MD calculations. We believe our theory provides a physically
realistic approach for the study of liquid dynamics.
\end{abstract}

\maketitle

A theory of the motion of atoms in a monatomic liquid has been developed, strongly motivated by the measured 
thermodynamic properties of elemental liquids. The potential energy surface underlying the motion is supposed
to consist of a very large number of intersecting nearly harmonic many particle valleys.    
Valleys whose structure (the configuration of atoms at the bottom of the valley) has crystalline or microcrystalline symmetry are 
relatively few in number, while valleys with the maximally disordered \textit{random} structures are of overwhelming numerical
superiority, and consequently dominate the statistical mechanics of the liquid.
 The identification of random valleys is an important step, since they all have identical macroscopic 
properties in the thermodynamic limit, hence have the same structural potential energy and the same 
distribution of normal vibrational modes. The corresponding atomic motion is comprised of two parts: 
normal mode vibrational motion within a single random valley, interspersed with motions across valley intersections,
called transits.  Further, the valley intersections are supposed to be sharp, and the transits nearly 
instantaneous, so that the equilibrium thermodynamic properties result from motion within a single random valley,
while the transits are responsible for diffusion. The presence of inherent structures in the liquid was demonstrated 
by Stillinger and Weber \cite{SW82,S95}, while the remaining details regarding the potential energy surface 
and the nature of the vibrational and transit motion were postulated in \cite{DW1}. The theory
has been confirmed by comparison with thermodynamic properties of elemental liquids \cite {DW1,DWcv}, by 
computer studies of random valleys and their normal vibrational modes \cite {Brad1, Brad2}, by observation of 
single transits in molecular dynamics (MD) calculations \cite{transits} and by a study of the velocity autocorrelation 
function and self-diffusion \cite{EricZ(t)}. A review has been presented \cite{Ericrev}.

Here we apply our theory to the analysis of inelastic neutron scattering in a monatomic liquid. We work with a small system, 
having $N=500$ atoms in a cubic box with periodic boundary conditions, and with density and interatomic
potential representing liquid sodium at melt.  Fourier components are evaluated at $\bm{q}$ consistent 
with periodic boundary conditions. Statistical averages are denoted $\langle \dots \rangle$, and consist of an
average over the atomic motion, in classical statistical mechanics, followed by an average
over $\bm{q}$ in a star, denoted $\langle \dots \rangle_{\bm{q}^{\ast}}$. Under these 
conditions it is appropriate  to compare theoretical calculations directly with MD, since finite-$N$ effects are the 
same in both, and the MD ensemble correction vanishes for fluctuations considered here \cite{DWbook2}. The same 
approach of comparing theory with MD was successfully used to evaluate the contribution of anharmoniciy 
to the dynamic structure factor of crystalline potassium \cite{GHK}.

The intermediate scattering function is $F(q,t)$,
\begin{equation}\label{F(q,t)}
F(q,t) = \frac {1}{N} \left < \rho (\bm{q},t) \rho (-\bm{q},0) \right >
\end{equation}
where $\rho (\bm{q},t) $ is the Fourier transform of the density operator,
\begin{equation}
\rho (\bm{q},t) = \sum_{K} e^{-i \bm{q}\cdot \bm{r}_{K}(t)},
\end{equation}
and the atomic positions are $\bm{r}_{K}(t)$ for $K=1, \dots,N$. When the system moves in a single random valley,
each particle has fixed equilibrium position $\bm{R}_{K}$ and displacement $\bm{u}_{K}(t)$, so that
\begin{equation}
\bm{r}_{K}(t) = \bm{R}_{K} + \bm{u}_{K}(t).
\end{equation}
When the displacements undergo harmonic vibrations, the average over the atomic motion is denoted 
$\langle \dots \rangle_{h}$, and the average in Eq.~(\ref{F(q,t)}) is $\langle\langle\dots\rangle_{h}\rangle_{\bm{q}^{\ast}}$.
Then for harmonic motion, $F(q,t)$ becomes 
\begin{widetext}
\begin{equation} \label{Fqtt}
F(q,t) = F(q,\infty)  + \frac {1}{N} \left < \sum_{KL} e^{-i \bm{q}\cdot \bm{R}_{KL}} e^{-W_{K}(\bm{q})} e^{-W_{L}(\bm{q})}
          \left [ e^{\left < \bm{q} \cdot \bm{u}_{K}(t) \; \bm{q} \cdot \bm{u}_{L}(0)\right >_{h}} -1 \right ]
           \right >_{\bm{q}^{\ast}},\\
\end{equation}
\end{widetext}
\begin{equation} \label{Finf}
F(q,\infty) =   \frac{1}{N} \left < \sum_{KL} e^{-i \bm{q}\cdot \bm{R}_{KL}} e^{-W_{K}(\bm{q})} e^{-W_{L}(\bm{q})}
\right >_{\bm{q}^{\ast}},                   
\end{equation}
where $\bm{R}_{KL}=\bm{R}_{K}-\bm{R}_{L}$, and where $W_{K}(\bm{q})$ is the Debye-Waller factor 
$\frac{1}{2}\langle(\bm{q}\cdot\bm{u}_{K})^{2}
 \rangle_{h}$. Notice the functions 
 $\langle\bm{q}\cdot\bm{u}_{K}(t)\;~\bm{q}\cdot\bm{u}_{L}(0)\rangle_{h}$ vanish as $t\rightarrow\infty$.

The dynamic structure factor $S(q,\omega)$ is the Fourier transform of $F(q,t)$. This can be evaluated analytically
in the one-mode approximation, obtained by expanding $e^{\langle \dots \rangle}-1$ to linear order in Eq.~(\ref{Fqtt}).
The transformation from $u_{Ki}$,where $i=x,y,z$, to the normal mode amplitudes $q_{\lambda}(t)$, where
$\lambda=1,\dots,3N$, is real orthogonal, with eigenvector components $w_{Ki,\lambda}$ and normal mode frequencies
$\omega_{\lambda}$. Then in the harmonic one-mode approximation, where the inelastic part of $S(q,\omega)$
is denoted $S_{1}(q,\omega)$, one has in classical statistics
\begin{equation} \label{S}
S(q,\omega) = F(q,\infty)\delta(\omega)+S_{1}(q,\omega),
\end{equation}
where
\begin{equation} \label{S1}
S_{1}(q,\omega)=\frac{3kT}{2M}\frac{1}{3N}\sum_{\lambda}f_{\lambda}(q)[\delta(\omega+\omega_{\lambda})+
\delta(\omega-\omega_{\lambda})], \\
\end{equation}
and the key function in $S_{1}(q,\omega)$ is
\begin{equation} \label{flambda}
f_{\lambda}(q)=\frac{1}{\omega_{\lambda}^{2}}\left < \left|\sum_{K}e^{-i\bm{q}\cdot\bm{R}_{K}}e^{-W_{K}(\bm{q})}
\bm{q}\cdot\bm{w}_{K\lambda}\right|^{2}\right >_{\bm{q}^{\ast}}.\\
\end{equation}
Apart from the thermal factor $3kT/2M$ and the normalization $1/3N$, $f_{\lambda}\delta(\omega+\omega_{\lambda})$ is the
cross section for neutron scattering at momentum transfer $\hbar q$ with annihilation of energy 
$\hbar\omega_{\lambda}$ in mode $\lambda$, and $f_{\lambda}\delta(\omega-\omega_{\lambda})$ is the cross section
for neutron scattering at momentum transfer $\hbar q$ with creation of energy 
$\hbar\omega_{\lambda}$ in mode $\lambda$. The normal mode resolution of the Debye-Waller factor is
\begin{equation} \label{WK}
W_{K}(\bm{q})=\sum_{\lambda}\frac{kT(\bm{q}\cdot\bm{w}_{K\lambda})^{2}}{2M\omega_{\lambda}^{2}}.
\end{equation}

\begin{figure}[h]
\includegraphics[height=3.0in,width=3.0in]{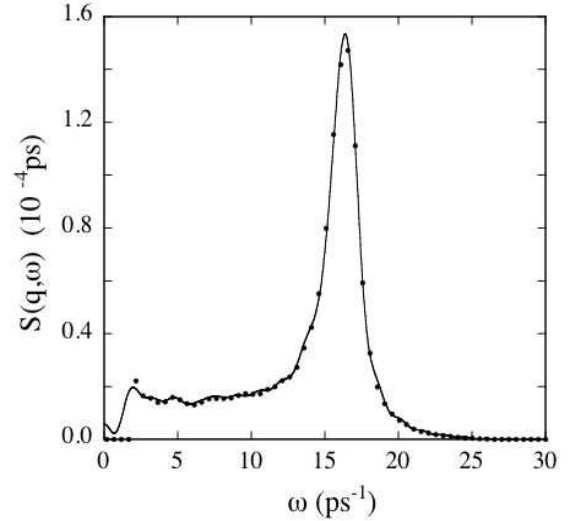}
\caption{\label{Sqw_T3} $S(q,\omega)$ for motion in a single random valley at $q=0.29711$ a$_{0}^{-1}$ and 24~K: full line is MD,
  points  are theory (Eq.~(\ref{S1})).}
\end{figure}

To evaluate $S(q,\omega)$ in MD, $F(q,t)$ was calculated directly from Eq.~(\ref{F(q,t)}), with $\langle\dots\rangle$
being the average over a long MD trajectory, followed by the average over the star of $\bm{q}$. Then the long-time
constant value of $F(q,t)$ was subtracted and the result, $F(q,t)-F(q,\infty)$, was Fourier transformed.
It is noteworthy that the theoretical harmonic $F(q,\infty)$ is in good agreement with the MD result.
Indeed, a small anharmonic contribution is also present, which brings theory and MD into perfect agreement
for $F(q,\infty)$. Further discussion of anharmonicity is omitted here, because it has negligible effect on
$S_{1}(q,\omega)$. To study the motion in a single random valley, the MD calculations were done for a system at 24~K, 
where no transits occur. The representative $\bm{q}$ is $\frac{2\pi}{L}(1,1,2)$, giving $q=0.29711$ a$_{0}^{-1}$, about
one third of $q$ at the first peak in $S(q)$. The result is shown in Fig.~\ref{Sqw_T3}.

Numerical evaluation of Eq.~(\ref{S1}) for $S_{1}(q,\omega)$ was accomplished by making a histogram, thus 
smoothing the $\delta$-functions. The $f_{\lambda}(q)$ were calculated for the same random valley that the
MD system was in, and at the same temperature, and the histogram is shown by the points in Fig.~\ref{Sqw_T3}.
 The contribution $F(q,\infty)\delta(\omega)$ is not 
indicated. For motion in a single random valley, the agreement between harmonic
one-mode theory and MD is highly accurate for $S(q,\omega$), and the same is true for all other $q$ values we studied.

\begin{figure}[h]
\includegraphics[height=3.0in,width=3.0in]{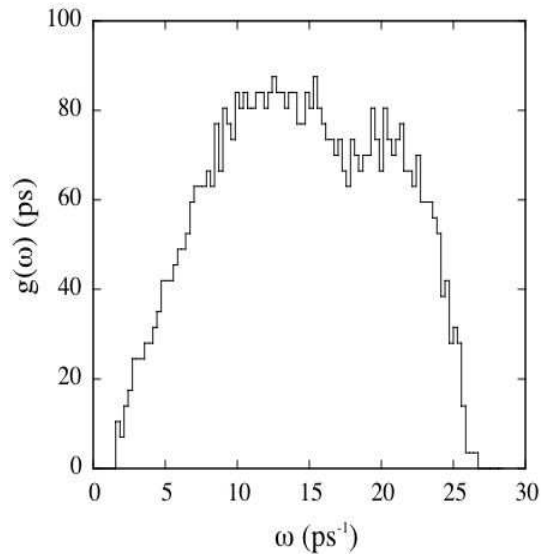}
\caption {\label{gw} Vibrational mode frequency distribution for a single random valley (1497 modes). }
\end{figure}

To assist in the interpretation of our results, the normal mode frequency distribution $g(\omega)$ is shown
in Fig.~\ref{gw}. Though the graph is for a single random valley, it is essentially the same for all random valleys.
Our system has no modes with frequency below 1.7 ps$^{-1}$; modes with lower frequencies will appear if the system size
is increased. The absence of such  modes accounts for $S_{1}(q,\omega)=0$ at low frequencies in Fig.~\ref{Sqw_T3}.

\begin{figure}[t]
\includegraphics[height=3.0in,width=3.0in]{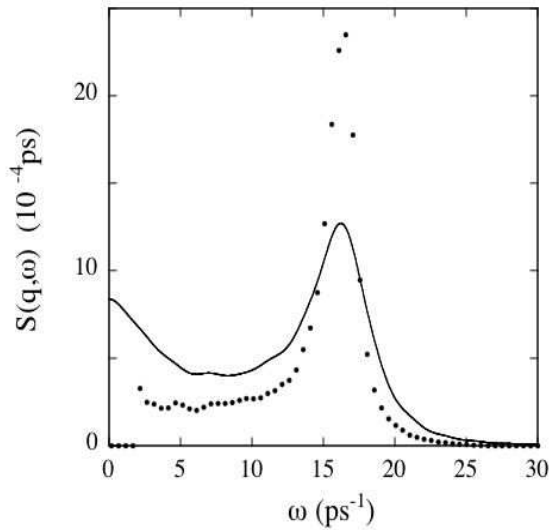}
\caption {\label{Sqw_T4}  $S(q,\omega)$ at $q=0.29711$ a$_{0}^{-1}$ and  395~K: line is MD for the liquid, 
points are theory for a single random valley.}
\end{figure}

The same information shown in Fig.~\ref{Sqw_T3}, but now at 395~K, some 24~K above melting, is shown
in Fig.~\ref{Sqw_T4}. The theoretical $S(q,\omega)$ still represents motion in a single random valley, so the
Rayleigh peak is $F(q,\infty)\delta(\omega)$ at 395~K, and the Brillouin peak has nearly the same location and width
as in Fig.~\ref{Sqw_T3}. On the other hand, the MD system at 395~K transits between random valleys at a very
high rate, producing two net effects in the Rayleigh and Brillouin peaks relative to their shapes in a single random 
valley: (a) the peaks are not shifted, and (b) each peak retains approximately the same area but is broadened. 
Point (a) needs to be understood first, so we shall address it here.

First, when transits are present, $\bm{R}_{KL} = \bm{R}_{K} - \bm{R}_{L}$ in Eqs.~(\ref{Fqtt}) and (\ref{Finf}) becomes
$\bm{R}_{K}(t) - \bm{R}_{L}(0)$. If now atom $K$ is involved in a transit from valley $a$ to valley $b$
at  time $t_1$, its motion in first approximation is as follows \cite{EricZ(t),Ericrev}: the atom is in vibrational
motion in valley $a$ before $t_1$ and  in valley $b$ after $t_1$; its position and velocity are continuous at $t_1$
but its equilibrium position $\bm{R}_{K}$ and its displacement $\bm{u}_{K}$ are discontinuous at $t_1$.
According to Eqs.~(\ref{Fqtt}) and (\ref{Finf}), the discontinuities in  $\bm{R}_{K}$ will disrupt correlations inside
the $\sum_{KL}$, and so will broaden the Rayleigh and Brillouin peaks; the discontinuities in $\bm{u}_{K}$
will hasten the decay of  $\left <\bm{q}\cdot\bm{u}_{K}(t)\;\bm{q}\cdot\bm{u}_{L}(0)\right >_{h}$ to zero, and 
so will broaden the Brillouin peak; but none of the discontinuities will cause the peaks to shift. The fact that 
the presence of transits does not cause the peaks  to shift in the MD system, according to Fig.~\ref{Sqw_T4}, is 
evidence that transits are indeed nearly instantaneous in the liquid.

\begin{figure}[h]
\includegraphics[height=3.0in,width=3.0in]{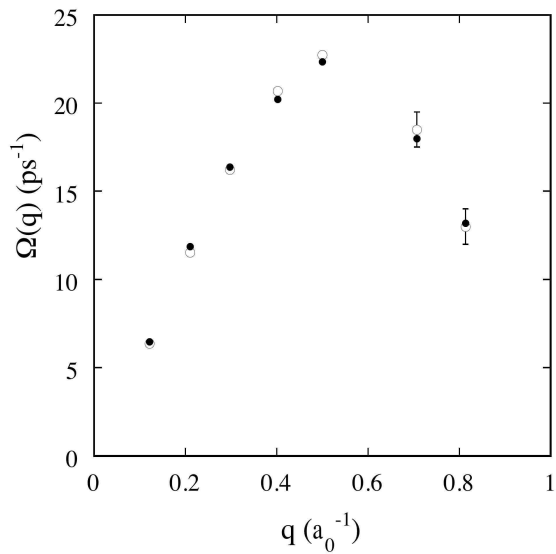}
\caption {\label{wq}  Brillouin peak maximum $\Omega (q)$ at 395~K: circles are MD for the liquid, points
   are theory for a single random valley. }
\end{figure}

It is well known that the Brillouin peak varies with $q$, and its maximum frequency $\Omega(q)$ follows a characteristic
curve \cite{dispcurve,Anees,ValB47,HMCD}. The question is, does the agreement shown in Fig.~\ref{Sqw_T4} 
hold for all
$q$? That it does is shown in Fig.~\ref{wq}, for the complete range of $q$, from the smallest allowed $q$ for
our system up to $q$ around 0.8~a$_{0}^{-1}$, where the Brillouin peak disappears. Analysis of our data for 
$\Omega(q)$ from theory for a single random valley, and from MD calculations, reveals that the theory and MD agree
within errors  at 24~K and at 395~K, with an overall rms relative difference of $0.02$ (negligible), and that 
the shift from 24~K to 395~K is zero within errors for theory and for MD,
with an overall rms relative shift of $0.02$ (negligible).

Our theory provides an interpretation of the $\Omega(q)$ dispersion curve which, as far as we know, is new: 
the maximum frequency in the dispersion curve is near the maximum frequency of the normal mode distribution. 
The maximum $\Omega$ in Fig.~\ref{wq} is around 23~ps$^{-1}$, while the distribution of Fig.~\ref{gw} 
has very few modes beyond this frequency. Hence liquid experiments can provide information on the random
valley vibrational mode distribution.  We note Fig.~\ref{wq} is in good agreement with experiments for liquid sodium
at 390 K \cite{Naexp}.

In the past twenty years or so, the dynamic structure factor has been extensively studied by means of mode
coupling theory \cite{GoLu75,Sjog80,GoSj92}. Recently, highly accurate inelastic x-ray scattering measurements have been
done for a wide range of $q$ (e.g. \cite{R&c002}). These data, and MD
 data as well,  have been analyzed by using an ansatz for the memory function 
  in the generalized Langevin equation for $F(q,t)$, in which the memory function possesses two (or sometimes three)
  decay channels for density fluctuations, with each decay process represented by adjustable $q$-dependent
  parameters for the decay strength and relaxation time \cite{R&c001,SBRS,SRSV}.  
  Though the fits to $S(q,\omega)$ data
  are excellent, it remains to interpret their meaning by identifying the physical process responsible
  for each decay channel \cite{R&c001,SBRS,SRSV}.

  In comparison, our theory is based on well defined physical processes from the outset, namely nearly 
  harmonic motion in a single random valley, plus nearly instantaneous transits.  Harmonic motion in a single
  random valley has been shown to account for the thermodynamic properties of elemental liquids, with small
   corrections from transits \cite{DW1,DWcv}. Here this harmonic motion accurately reproduces MD for $S(q,\omega)$
   in the glass state, Fig.~\ref{Sqw_T3}, and  this motion also accounts for the  Brillouin 
   peak dispersion curve in the liquid, where the system moves rapidly among a host of
   random valleys, as shown in Fig.~\ref{wq}. Moreover, the interpretation of the 
   vibrational contribution to the Brillouin peak is clear from Eq.~(\ref{S1}): the peak at a fixed $q$ is
   the sum of cross sections from a band of normal modes which scatter efficiently
   at that $q$. Transits, on the other hand, are the liquid counterpart of binary collisions in a gas:
   a transit is the highly correlated motion of a small local group of atoms that carries the system across
    the boundary between two random valleys. Single transits have been resolved in MD calculations at low
    temperatures \cite{transits}, and a model for transits gives a good account of the velocity autocorrelation 
    function from the glass state to the very hot liquid \cite{EricZ(t)}. While much study 
   of the transit process remains, transits are nevertheless well defined, and their role of broadening the peaks of
   $S(q,\omega$) in the liquid state is expressed through Eqs.~(\ref{Fqtt}) and (\ref{Finf}) for $F(q,t)$  as 
   discussed in the text. In summary, we believe the case has been made that our representation of the atomic motion 
   provides a physically realistic theory of liquid dynamics.
   
   The authors gratefully acknowledge Eric Chisolm and Brad Clements for collaboration, and Francesco Venneri for technical 
   advice. This work was supported by the US DOE through contract W-7405-ENG-36.

\bibliography{INS}

\end{document}